# Designing Beyond Current Conceptualizations of Spaceflight Experiences

James Cole
School of Behavioral and
Brain Sciences
The University of Texas at Dallas
Dallas, TX, USA
james@jamescole.info

Kathryn Hays
Department of Information
University of North Texas
Denton, TX, USA
Kathryn.Hays@unt.edu

Ruth West
Department of Design
University of North Texas
Denton, TX, USA
Ruth.West@unt.edu

## ABSTRACT

The potential future democratization of spaceflight reveals a need for design of experiences that extend beyond our current conceptualization of spaceflight. Research on career astronauts indicates that transformative experiences occur during spaceflight despite the physiological and psychological stressors involved. This phenomenon allows us to envision a future where such profound experiences are accessible to diverse spaceflight participants. In this position paper, we advocate for acknowledging how design decisions made at the genesis of commercial spaceflight might impact space travelers of this speculative future. In proposing salutogenesis as an orienting topic, a potential design framework, and as a metric for spaceflight participant experience, we offer a call to action for the broader experience design community to engage with the design of profound experiences for spaceflight participants.

## CCS CONCEPTS

• Human-centered computing → Human computer interaction (HCI)

## KEYWORDS

Commercial space tourism, Spaceflight participant, Passenger experience, Human-Centered design

## 1 SPACEFLIGHT AND A DIVERSE PUBLIC

Expansion of the commercial spaceflight industry is accelerating, with highly publicized flights from operators such as SpaceX, Blue Origin [24], and Virgin Galactic (that identifies itself as the "world's first commercial spaceline" with the purpose of "connecting people across the globe… to space travel") [25]. Blue Origin flight participants span tech moguls, prior space exploration personnel from the International Space Station (ISS), and celebrities [14]. Acknowledging this acceleration and pointing towards a future democratization of spaceflight, the FAA no longer designates individuals as astronauts [23] and instead offers recognition of human spaceflight participants [26]. While potentially a distant and speculative future, these flights capture the popular imagination and envision the possibility for members of a broad and diverse public to experience spaceflight. This reveals a need to develop approaches for designing experiences that would take place both beyond our lifetime and current conceptualization of spaceflight. Design of commercial spaceflight for a diverse public may require the development of equally diverse methods. Our work asks: *What kinds of experience design approaches can inform the future development of commercial spaceflight for a broad and diverse public?*

Researchers and experience designers utilize a range of methods to understand and design for real world and virtual experiences, including spaceflight contexts. Florom-Smith [4:2] notes that the proprietary nature of psychological assessments and training for commercial spaceflight operators may prevent accurate gauges of commercial spaceflight participant (SFP) safety. Applying experience design methods to the design of future commercial spaceflight requires the development of open, data-driven approaches that parallel the methods derived from studies of career astronauts. While research conducted with astronauts provides an understanding of design for physiological and psychological well-being during spaceflight, this knowledge may not fully generalize to the design of future spaceflight experiences driven by SFPs' goals, such as the opportunity to experience awe and the ability to "look down at the Earth from space" [11:45] or goals as yet to be envisioned.

We propose steps towards experience design for the comfort and care of commercial spaceflight participants, not only the mitigation of physiological and psychological risk, but also an understanding of the human experience *writ large* in these future contexts. We propose the concept of salutogenesis as an orienting topic, a potential design framework, and as a metric for a holistic SFP experience, with the potential to complement existing methods that optimize safety and wellness. Below, we offer a call to action for the broader experience design community to engage with the design of profound experiences for SFPs.

## 2 DESIGNING FOR A NEW KIND OF SPACE TRAVELER

Early human spaceflight missions (1961–1970) were relatively short in length, limited to only a few days, which minimized the impact of adverse psychological conditions [6:5]. From 1971–1984, missions that only lasted a few weeks began to extend to several months. In 1987, the first mission lasting one year was conducted successfully. As long-duration missions became more commonplace, including those aboard the ISS, the need for behavioral health support for astronauts became increasingly apparent [15:124]. Today, behavioral health programs, such as NASA's Behavioral Health and Performance Group, draw from more than 60 years of human spaceflight experience to provide psychological support for astronauts as they adapt to the stressors of living and working in space [27]. Among the psychological stressors characteristic of spaceflight are isolation, confinement, danger, monotony, and workload [6:2].

Since existing research on the psychology of spaceflight has been conducted on homogeneous groups of highly-trained astronauts, little is known about the psychological effects of spaceflight on SFPs [12:882,16:1053]. Maximizing the safety and enjoyment of SFPs will require the creation of psychological evaluations and training protocols specific to commercial space vehicles and experiences, but no plans of this sort are currently published [4:6–7]. Ligor et al. [8:2] notes how treating existing safety practices as proprietary information limits the development of consensus standards for commercial human spaceflight, and stresses the importance of collaboration between government, academia, and industry in the development of standards. Since 1998, cooperation of this kind has occurred between the five space agencies represented on the ISS in their effort to implement standard behavioral health programs for ISS crews [15:122–123]. Similar policies of openness and collaboration within the commercial spaceflight sector may enable the creation of standard tools and methods for responding to adverse psychological states in SFPs.

The prevalence of anxiety related to air travel, such as fear of flying (FOF), is around 20% within the general population [12:883]. Despite the likelihood that conditions like FOF will impact space operations, the FAA offers no specific recommendations for protecting SFP psychological health beyond keeping "stress at levels that can be considered safe for space flight participants" [28:4]. Designing experiences for SFPs requires a deeper understanding of the psychological stressors of spaceflight that are unique to this demographic. FOF and centrifugation studies demonstrate how simulations can be used to elicit stressors analogous to those encountered during spaceflight [12:882]. However, as of 2019, few studies investigated the effects of centrifugation on SFPs [16:1057], limiting the generalizability of the findings. The expansion of spaceflight to a diverse public makes investigations into the psychological stressors and formulation of countermeasures paramount to the design of optimal experiences [4:7]. Human-computer interaction (HCI) researchers and user experience (UX) designers, trained to apply insights of human behavior in the design of technology, could be ideal collaborators in these investigations.

## 3 BEYOND EXISTING DESIGN APPROACHES

The integration of psychology and interaction design has an established history in aerospace applications. During World War II, work from human factors pioneers, like Paul Fitts and Alphonse Chapanis, to reduce the number of plane crashes through the design of more usable aircraft controls highlights the significance of mapping designs to human needs [7:80–87]. Since then, governmental space programs like NASA have made significant strides in understanding and securing the physiological safety of humans outside the bounds of Earth's atmosphere. Beginning at the conceptual phase of system design, NASA's Human Systems Integration (HSI) practitioners factor human capabilities, limitations, roles, and responsibilities into the design of hardware and software systems [19:1–1,20:21].

HCI researchers and UX designers employ various design frameworks to center user needs within the design process so that action can be oriented towards addressing specific needs from the start. Design frameworks, like the Double Diamond [21] and Design Thinking [3], enable structured approaches for solving design challenges. They require that problems be defined in specific enough terms so that actions can be taken to solve them. Frameworks are not intended to be limiting or prescriptive, but instead offer ways to visualize how decisions made throughout the design process affect the outcome [17:126]. The four-phase Double Diamond process from the British Design Council [21] identifies user needs during the *discovery* phase. Similarly, the Design Thinking framework addresses user needs within the *empathize* phase, during which designers categorize the needs identified during user research along a matrix of what the user says, thinks, feels, and does [3]. Human-centered design, with a history of use at NASA [19:20], represents yet another method for incorporating user perspectives into the design process to develop effective systems [22]. Each of these processes requires an understanding of the contexts in which specific users interact with a design, incorporates users into the design process, develops prototypes, and refines the design through testing and iteration.

Commercial spaceflight operators are already applying user-centered design methods to the creation of user interfaces for crew controls within vehicles. For example, the touchscreen user interfaces found in the SpaceX Crew Dragon spacecraft were validated by user testing with a diverse set of participants. These efforts enabled the design of UI controls that account for crew members' gloved hands and words that are legible despite the high vibrations users would encounter during flight [9]. The design of systems intended for passenger interaction should be informed by similar observations of user needs and behaviors contextualized within the SFP experience.

Bernard et al. [2] proposes the people, environment, actions, and resources (PEAR) model to identify the human factors of passenger experience that designers should consider in regards to spaceflight contexts. The PEAR model enables the design to center on specific details which are important to each stakeholder in a commercial spaceflight experience, crew and passengers alike. Addressing the future of space tourism and SFPs, Bernard acknowledges that work is needed to go beyond existing design approaches, and actively engage human factors and experience design teams in developing the nascent space tourism industry. Here we concur, researchers working with user experience designers are well positioned to apply insights of human behavior in the design of commercial space flight experiences. Extending the focus of research and design beyond risk mitigation and safety concerns enables interdisciplinary teams to consider the comfort and care of SFPs and the potential to design profound experiences. Below we discuss one possible avenue for this engagement.

## 4  SALUTOGENESIS: A DRIVER FOR EXPERIENCE DESIGN

Astronauts perform mission-critical tasks while simultaneously adapting to the stressors of space. Any cognitive performance decrements, including those caused by diminished psychological well-being, pose risks to both safety and mission success [6:2–3]. Psychological assessments and training protocols, designed with data collected throughout the course of human spaceflight programs, prepare astronauts to adapt to the extreme environments in which they operate. Similarly, identification of the psychological factors specific to commercial spaceflight participation will enable the development of innovative screening tools, training protocols, and in-flight technologies geared towards reducing the impact of negative psychological events during spaceflight for SFPs. However, operational models designed around the needs of astronauts may not be entirely suited for the needs of SFPs. Furthermore, existing research on spaceflight psychology centers mostly on the risks of adverse psychological conditions and less is known about "the potential behavioral health benefits of space travel" [6:135,18:2]. Design approaches that emphasize risk mitigation to the exclusion of other psychologically salient phenomena may miss an opportunity to deliver holistic experiences that elevate the positive psychological aspects of human spaceflight.

Mittelmark & Bauer [10] assert that positive psychology and salutogenesis are conceptually consonant despite differences in terminology. Salutogenesis embodies a continuum of well-being [1:12–37], with a focus on the enhancement of health and well-being through a sense of coherence between stressors and one's lived experience [5]. Powerful and challenging experiences can enhance or bring about well-being [6:135]. Coherence emerges as the integration of comprehensibility, manageability, and meaningfulness of stressors gives rise to the transformative experience of salutogenesis [10].

Ritsher et al. [13] highlights how salutogenesis resulting from adaptation to the extreme environment of space can produce both temporary and lasting benefits. Post-mission surveys and other data collected from career astronauts indicate a universally positive reaction to being in space, noting changes such as "an enhanced appreciation for the beauty and fragility of Earth" [6:138]. The overview effect, an intense feeling experienced by viewing the Earth from above that is often reported by astronauts, is among the factors that may lead to salutogenesis [18:7]. Other positive factors associated with the environmental and social contexts of spaceflight converge to generate these profound experiences, such as membership in elite groups, changed values, and self-confidence [15:121–122]. Maximizing the potential for salutogenesis can offset the risks posed by negative psychological events that manifest during spaceflight [13:1].

Existing approaches to passenger experience hint at the attention that may be directed towards positive emotional considerations as commercial human spaceflight evolves. For instance, Virgin Galactic's marketing materials describe their hope that experiencing the overview effect will intensify a passenger's "connection to Earth and to humanity" [29:1, 13]. Using salutogenesis as a metric in experience design could offer a benchmark for evaluating the affective qualities of spaceflight experiences more comprehensively. HCI professionals could collaborate with industry, academia, and multidisciplinary entities to define both the negative and positive psychological aspects of commercial spaceflight experiences. Individual goals, like producing the overview effect, could be included in this holistic interpretation of passenger experience. This can inform the creation of training protocols and simulations optimized for the needs of SFPs, enable the development of technological countermeasures for responding to adverse psychological conditions, and maximize the potential for generating salutogenic experiences during spaceflight.

The inherently human yet transformative nature of salutogenic experiences makes them candidates for accessing the underlying components of speculative and envisioned future experiences of spaceflight. In this paper we advocate for a discourse towards salutogenic experience design for SFPs. Focusing design efforts on the factors that lead to salutogenesis could serve to both lessen the impact of adverse psychological conditions during spaceflight and enable the design of profound experiences for SFPs.

## 5  CALL TO ACTION

Although this paper describes a speculative future, the democratization of space demands that we acknowledge how design decisions made at the genesis of this industry impact people who do not yet have access or who may not yet exist. How are we as researchers and user experience designers to design commercial spaceflight as a holistic experience? If what we know about designing for the care and comfort of spaceflight participants comes from research on career astronauts, or proprietary information with limited access, how are we to have a broader dialogue inclusive of the diverse voices of those who may participate in the future of spaceflight?

The ideas presented here are intended to provide a starting point. They encourage the broader experience design community to engage with the design of profound experiences for SFPs. They propose salutogenesis as an orienting topic, a potential design framework, and as a metric for SFP experiences, complementing existing methods used to optimize human safety and expanding focus from solely risk mitigation towards transformative and profound spaceflight participant experiences.

## ACKNOWLEDGMENTS

We thank Christopher Lueg for his generous comments on drafts of this paper.